# Design and performance of a Toroidal RF Volume Coil with Intrinsic Electromagnetic Interference Rejection for low-field Portable Halbach-Based MRI Systems

Jules Vliem[1], Najac Chloe[2], Beatrice Lena[2], Andrew Webb[2], Irena Zivkovic[1*]

[1]Electrical Engineering Department, Eindhoven University of Technology, Eindhoven, The Netherlands

[2]C.J. Gorter MRI Center, Department of Radiology, Leiden University Medical Center, Leiden, The Netherlands

[*]Corresponding author: i.zivkovic@tue.nl

## Abstract

**Purpose:** One of the intrinsic limitations of low-field MRI is low signal-to-noise ratio (SNR), which can be further reduced by electromagnetic interference (EMI) due to the lack of a Faraday shielded room. To address this issue, we propose a novel RF coil design that is inherently less sensitive to EMI while maintaining high receive sensitivity.

**Methods:** The proposed coil structure is based on an anapole (toroidal) design and consists of six rings, each containing four continuous wires wound around an elliptical 3D-printed former. The wire in the entire structure is uninterrupted. This coil is designed for use in systems with an axial $B_0$ field direction. The performance of the proposed RF coil was evaluated on a Halbach array system designed for neuroimaging and operating at 47mT and compared with a widely used spiral head coil.

**Results:** The noise level achieved by the proposed toroidal coil in combination with a belt wrapped around the subjects and grounded to the scanner was comparable to that of the widely used spiral head coil when combined with a grounding belt and passive aluminum shielding. The coil transmit and receive efficiency was comparable to the efficiency of the spiral head coil.

**Conclusion:** The proposed RF coil inherently reduces the effect of EMI, potentially removing or reducing the need for passive shielding or external/internal sensors for EMI reduction in post-processing.

# 1. Introduction

Low-field MRI offers a cost-effective imaging solution accessible to a wide population, particularly in terms of the development of lightweight, portable scanners[1–4]. However, a fundamental limitation of low-field systems is their inherently low signal-to-noise ratio (SNR)[5]. Addressing this limitation is essential to improving the utility of low-field MRI. As reported in [6,7], both hardware and software enhancements can contribute to SNR improvements.

One of the major sources of noise in low-field MRI systems is electromagnetic interference (EMI), which can originate from internal components such as RF and gradient electronics, as well as from external environmental sources. EMI induces noise in the receive coils, compromising image quality, and thus its mitigation is critical for the performance of low-field MRI. The body acts as a very effective antenna at low frequencies, which further increases the effect of EMI [8].

In conventional high-field systems, EMI is typically managed using enclosed RF shielding rooms (Faraday cages). However, this solution is not practical for portable, bedside-compatible low-field scanners, which prioritize a compact form factor and minimal infrastructure.

To address EMI in low-field systems, recent approaches have involved deploying multiple RF coils both inside and outside the scanner [9–12]. These coils must be synchronized with the receiver chain for effective noise cancellation. Nonetheless, using RF coils for EMI cancellation introduces challenges. In [9], external receive coils were employed to reduce EMI, but the resulting SNR was less than half that of configurations with flexible shielding. To overcome this, a method described in [11] combines strategically positioned sensors with deep learning algorithms to more effectively model and subtract EMI, demonstrating a promising direction for future low-field MRI noise reduction strategies. A limited set of detectors is insufficient to capture the diverse EMI spectrum encountered in real-world environments. Optimal performance requires a large number of sensors, leading to increased system complexity and higher costs.

An alternative single-coil EMI suppression strategy was proposed in [12], using a passive power splitter/combiner in conjunction with a circularly polarized RF coil. In this configuration, one port is dedicated to EMI detection, while the other handles both MRI signal acquisition and EMI detection. The EMI signal can then be subtracted directly into hardware; however, this method also has an SNR penalty in that a birdcage geometry was used, which is inherently less sensitive than a solenoid.

To reduce or even eliminate the need for external shielding and RF sensors in mitigating EMI, we sought to design a volume RF coil with high transmit/receive sensitivity but being insensitive to external EMI. In antenna theory, such a structure is referred to as an "anapole" or non-radiating antenna [13–18], characterized by inherently complex geometries. The anapole concept was first introduced by Zeldovich in 1957 [13], who noted that the simplest form of an anapole antenna is a thin solenoid shaped into a toroidal configuration. Building upon this idea, we designed a volume coil composed of interconnected toroidal rings. We began by investigating how various parameters of the toroidal coil structure—such as twisting density, holder thickness, and the number of wire windings per holder—affect the coil's performance. After synthesizing the complete structure, we evaluated its performance through both simulated and experimental transmit efficiency and compared the results with those of a commonly used spiral head coil [4]. Finally, we assessed the efficacy of the toroidal coil, in combination with the grounding belt, to reject noise in vivo, and compared the results to the setup with the spiral head coil, in conjunction with the grounding belt and aluminum shield.

## 2. Methods

### 2.1 Electromagnetic simulations

Electromagnetic simulations were performed in CST Studio Suite 2025 (Dassault Systèmes, France). The toroidal coil geometry was designed in SolidWorks 2025 (Dassault Systèmes, France) with an elliptical shape measuring 180 mm in width and 240 mm in height (Figure 1a). These dimensions were chosen to match the reference spiral head coil design [4]. Note that these dimensions are with respect to the center of the coil holder; i.e. with a 10 mm thick holder, the resulting inner dimensions are 170 mm × 230 mm. The coil windings were modeled as 1.5 mm diameter copper wires twisted around the plastic holder (Polylactic acid (PLA), $\varepsilon_r$ = 3.2). A small separation of 0.5mm was maintained between the wire and the holder to avoid geometric intersections. All coils were placed in a cylindrical copper shield with an inner diameter of 305 mm, a length of 350 mm and a thickness of 1.25 mm. Simulations were conducted using a tissue-mimicking homogeneous phantom (width = 150 mm, height = 200 mm, $\varepsilon_r$ =83.4, $\sigma$ = 0.55 S/m). The frequency-domain solver with tetrahedral meshing was used, with open boundary conditions in all directions. Results were normalized to 1 W input power.

Initial simulations investigated the influence of the twisting density and holder thickness on transmit efficiency ($B_1^+$) at the center of the phantom. When varying the twisting density, the holder thickness was fixed at 10 mm; conversely, when varying the holder thickness, the twisting density was fixed at 24 twists per full winding.

After identifying an optimal configuration with six twists per winding and a 10 mm holder thickness, we evaluated various winding strategies/configurations. Figure 2 illustrates these different configurations. A simple elliptical copper wire served as a reference. Increasing the number of windings effectively resulted in multiple helical passes around the holder. For instance, the triple-wound toroid completes three full turns, with each turn offset by 120° relative to the others, resulting in a total of 18 twists across the structure. All coils were modeled using a single continuous conductor with a single feed gap.

The best-performing toroidal configuration was then extended into a six-element array with 25 mm center-to-center spacing, forming a 150 mm long volume RF coil (Figure 3). For comparison, simulations were also conducted on the reference spiral head coil, which was also 150 mm in length and consisted of 15 turns with 10 mm spacing between windings. Both the toroid and spiral head coils were capacitively segmented at their midpoint for tuning purposes.

## 2.2 Coil construction

The manufactured toroidal volume coil is shown in Figure 3. The holders and rings were 3D printed (Ultimaker S3, The Netherlands) using PLA filament (2.85 mm, Ultimaker). Each ring featured a groove to guide the wire and enforce a predefined twisting density and configuration. The rings were individually wound with Litz wire (1500 strands, 0.03 mm, Elektrisola, Germany) and subsequently soldered together to form a continuous structure.

A segmentation point was introduced between the third and fourth ring, where a tuning board was inserted. This board included a 150pF fixed capacitor (ATC 800E, USA) in parallel with a 1-30 pF variable capacitor (Johannson, USA) for fine-tuning. To enable easy tuning when a subject is inside the scanner, this tuning board was positioned outside of the main coil structure (Figure 3c).

At the end of the final ring, the wire was routed back towards the feeding point. The coil was tuned and matched to 1.98 MHz using a pi-matching network consisting of a 151 pF parallel capacitor and two 68 pF series capacitors. A schematic of the coil can be seen in Figure 3c.

A 3D printed base was designed to position the coil in the middle of the scanner. A small lid was put on top of the base so that the subject's head is not directly touching the structure. Additionally, a thin plastic sheet was placed inside the coil so that the patient didn't accidentally touch the wires.

A 15 turn, 240mm high, 180mm wide, 150mm long spiral head coil based on similar designs [4] in literature was designed and printed on a 3D printed former. The coil was wound with the same Litz wire as the toroidal coil and capacitively segmented halfway.

Both the toroidal and conventional spiral head coils were tuned and matched to 1.98 MHz, achieving a return loss ($S_{11}$) of less than −18 dB. Under in-vivo loaded conditions, the quality factors ($Q_{loaded}$) were 255 for the toroidal coil and 183 for the spiral head coil, respectively.

## 2.3 MR measurements

All measurements were performed on a 47 mT Halbach-based MRI system using a Magritek Kea2 spectrometer [2]. Comparisons were made between the toroidal coil and a standard spiral head coil.

### 2.3.1 $B_1^+$ Mapping

Transmit field ($B_1^+$) maps were measured on a head-shaped phantom filled with copper sulfate-doped water ($T_1 \approx 100$ ms). The phantom dimensions were 190 mm (anterior-posterior), 160 mm (left-right), and 240 mm (head-foot). $B_1^+$ mapping was performed using a 3D double-angle method [19][18], with two gradient echo (GRE) scans acquired at nominal flip angles of 60° and 120° (TR/TE = 600/12 ms, FOV = 200 × 220 × 190 mm³, resolution = 3 × 3 × 10 mm³, bandwidth = 20 kHz, RF pulse length = 200 µs, 1 average, scan time = 12 min 32 s per scan). K-space data were filtered with a sine-bell window (filter strength = 0.2) prior to reconstruction. Flip angle maps ($\alpha$) were computed using:

$$\alpha = \cos^{-1}\left(\frac{s_2}{2s_1}\right) = \tau \gamma B_1^+$$

where $s_1$ and $s_2$ are the signal intensities from the 60° and 120° scans, $\tau$ is the RF pulse length and $\gamma$ is the gyromagnetic ratio. For both coils during $B_1^+$ field measurements, the scanner bore was partially enclosed using a grounded copper plate to reduce external EMI. $B_1^+$ maps were normalized by the square root of the input power to estimate transmit efficiency. Background noise outside the phantom was removed via intensity thresholding.

## 2.4 EMI measurements

In unshielded environments, EMI can couple into the RF coil(s), with the human body further amplifying this effect by acting as antenna [8]. This issue can be mitigated by grounding the subject to a common ground shared with the scanner electronics [20]. In vivo experiments were performed using both coils under three different grounding conditions (Figure 5):

1. No grounding between the subject and scanner,
2. A grounding belt (in contact with the subject's lower back) connected to the scanner ground, and
3. The grounding belt plus two semi-cylindrical aluminum RF shields positioned laterally around the torso [6].

The grounding belt is made of a conductive polyester fabric metallized with copper and nickel (YSHIELD, Ruhstorf an der Rott, Germany) and is covered with a flexible, easy-to-clean plastic material for hygiene purposes (Kohlas, Corbas, France).

Images were acquired using a turbo spin echo (TSE) sequence (TR/TE = 600/16 ms, FOV = 205 × 225 × 190 mm³, resolution = 1.5 × 1.5 × 5 mm³, bandwidth = 20 kHz, RF pulse duration = 200 µs, echo train length = 8, scan time = 5 min 4 s) across all three conditions and for both coils.

Image SNR was estimated by placing a 15 × 15 pixel ROI in the background corners to measure the standard deviation of the noise, and an identically sized ROI in a central area of the brain to measure mean signal (Figure 5). The SNR was computed as the ratio of these two values.

In addition, noise profiles were acquired with a TSE sequence without RF excitation. A total of 1000 noise traces were recorded, which were then combined to estimate the noise levels under different grounding conditions. In total four volunteers were scanned.

## 3. Results

Figure 1 shows that increasing the number of twists per unit length leads to a reduction in the central $B_1^+$ field. A similar, though less pronounced, decrease is observed with an increase in holder thickness. A 10 mm thick holder was selected above a smaller diameter (e.g. 7.5 mm) to ensure mechanical stability. In Figure 2, it is evident that increasing the number of windings around a single ellipsoid enhances the $B_1^+$ field. Going from a single wound to a quadruple wound toroid resulted in a 45% increase in central $B_1^+$. Based on these observations, the final design was selected to include six twists per circumference and four full windings (quadruple-wound), with a holder thickness of 10 mm, as shown in Figure 3.

Figure 4 compares the simulated and measured $B_1^+$ fields of the toroidal and spiral head coils on a phantom. The simulation and experimental data show good agreement, with deviations in mean $B_1^+$ below 10%. In both cases, the toroidal coil exhibited approximately 10% higher transmit efficiency than the spiral head coil, though at the cost of increased field inhomogeneity. Simulated and experimental results showed that the toroidal coil has 3-3.5× higher relative inhomogeneity (coefficient of variation = standard deviation / mean) compared to the spiral head coil.

Figure 5 presents in vivo measurements using the toroidal and spiral head coils under three different conditions. Without the grounding belt and aluminum shield, both coils produced images with high noise levels. When the grounding belt was used, the toroidal coil achieved SNR more than five times higher than that of the spiral head coil under the same condition. When both the grounding belt and aluminum shield were applied, the toroid coil exhibited ~20% higher SNR compared to the spiral head coil. Corresponding noise peak measurements showed that the toroidal coil with only the grounding belt achieved noise suppression comparable to the spiral coil using both the grounding belt and aluminum shield. Similar noise levels were also observed when the toroidal coil was used with both the grounding belt and aluminum shield.

Figure 6 shows the repeatability measurement for both coils using only the grounding belt. Four volunteers were scanned with each coil, and the toroidal coil consistently delivered noticeably higher image quality across all subjects – with an average SNR of approximately 5 times greater than that of the spiral head coil.

## 4. Discussion

In this work, we proposed a novel RF volume coil design based on a toroidal structure for low-field MRI applications. The coil demonstrated slightly higher efficiency compared to the commonly used spiral head coil and exhibited more effective noise rejection. In in vivo measurements, the proposed coil—when used with only a grounding belt—achieved image quality comparable to that of the spiral head coil used in combination with both a grounding belt and an aluminum shield. Further optimization of the coil could potentially eliminate the need for bulky metallic shielding, thereby facilitating the use of low-field MRI systems as bedside instruments and in intensive care units.

When increasing the twisting density of a single-wound toroidal coil, a decrease in central $B_1^+$ was observed (Figure 1b). This effect can be attributed to a shift in the dominant current direction. As the wire is twisted more tightly around the holder, a larger fraction of the current flows axially (along the z-direction), generating fields orthogonal to the desired plane and thereby reducing the effective azimuthal current component responsible for producing $B_1^+$ in the desired transverse plane. A loosely twisted single wire produces the highest $B_1^+$ in the center, as it primarily carries azimuthally oriented current. However, if the twists become too loose, the toroid might lose its EMI rejection properties.

Configurations with lower twist density but multiple windings (e.g., quadruple-wound) maintain a dominant azimuthal current component while increasing the number of contributing loops, resulting in constructive reinforcement of the $B_1^+$ field along the z-direction. This behavior is analogous to a solenoid, where increasing the number of windings enhances the magnetic field strength by accumulating the contributions of each loop. Similarly, adding more windings in the toroidal design increases the $B_1^+$ efficiency.

While adding multiple windings to the toroidal structure can enhance the $B_1^+$ field, it also introduces several trade-offs. First, more windings increase the total wire length, and with it, the coil's resistance—leading to higher power dissipation and potential SNR degradation. To mitigate this, it is essential to use low-resistance conductors such as Litz wire. Second, the improvement in $B_1^+$ is not linear with added windings. For example, doubling the number of windings from the double-wound (17.21 µT/√W) to the quadruple-wound configuration (23.20 µT/√W) yielded only a 25% increase in $B_1^+$, despite doubling the wire length. Finally, practical limitations constrain how many windings can physically be implemented on the holder. It is still to be investigated if densely overlapping wires influence observed noise rejection capability of the proposed coil.

There is still room for improvement in the current design. Future iterations could explore the influence of the spacing between the rings (currently 25 mm), increasing the number of windings and a more detailed investigation into how the various toroidal parameters influence the noise rejection. Prior work [21] suggests that multiple segmentation of solenoid coils can reduce their sensitivity to load variations and external noise. We explored additional segmentation in our toroidal coil design via simulations, but did not observe a significant change in performance. Further investigation is needed to assess the role of segmentation in toroidal geometry.

An important advantage of the toroidal coil design lies in its intrinsic ability to suppress external EMI. The toroidal current distribution confines the magnetic field primarily within the toroid's core and near the windings, thereby reducing its interaction with external fields. Such a configuration approximates a non-radiating "anapole" mode, in which the electric and toroidal moment destructively interfere in the far field[13–17], minimizing radiation and susceptibility to external noise sources. While ideal anapole modes require perfectly symmetric current paths, the discrete and finite nature of real windings introduces some "leakage" fields. Nevertheless, the toroidal geometry offers a trade-off between $B_1^+$ performance and EMI rejection.

Future work will also explore the scalability of the proposed concept for imaging other anatomical regions, such as the extremities, while preserving its inherent noise rejection properties. Additional future work would include adaptation of this concept for operation in scanners where the $B_0$ field is aligned along the z-axis, as is typical in most clinical and UHF systems. If the coil's effectiveness proves to be independent of field strength, it could significantly simplify MRI system installations by eliminating the need for dedicated Faraday shielding.

## 5. Conclusion

We have presented a toroidal RF volume coil design tailored for low-field Halbach MRI systems, offering intrinsic EMI suppression. Our results demonstrate that this design can achieve image quality comparable to a conventional spiral head coil while requiring only a simple grounding belt and no additional shielding. By reducing the need for bulky shielding infrastructure, this approach simplifies scanner design, lowers cost, and enhances portability — making low-field MRI systems more suitable for use in settings such as intensive care units. Future work will explore scaling the concept for imaging other anatomies and assess its noise rejection performance across different magnetic field strengths.

**Acknowledgments**

Study conception: IZ; coil building and simulations: JV; acquisition of data: JV, CN, BL, AW, IZ; Funding acquisition: IZ; All authors have participated in the analysis and interpretation of data, manuscript drafting and critical revision.

JV was supported by the Dutch Growth Fund program POLARIS (www.polaris-ngf.nl).

**Figures:**

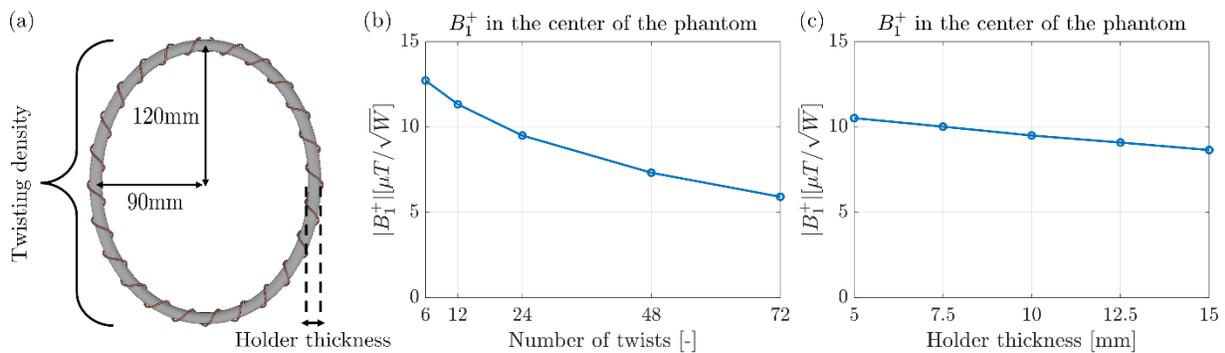

**Figure 1:** (a) Model of the single-wound, elliptical toroid, with dimensions indicated (from the center of the holder). (b) Influence of twisting density and (c) holder thickness on the simulated central $B_1^+$ magnitude in the phantom. Results are normalized to 1 W accepted power.

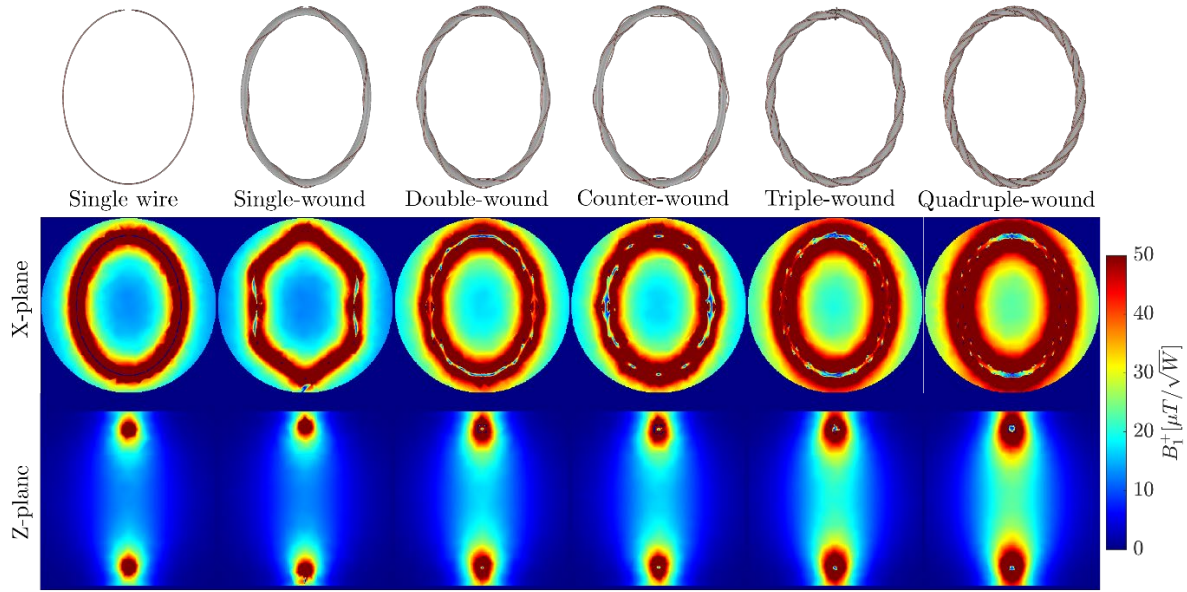

| Type | Single wire | Single-wound | Double-wound | Counter-wound | Triple-wound | Quadruple-wound |
|---|---|---|---|---|---|---|
| Impedance [$\Omega$] | 0.05+j7.0 | 0.05+j7.4 | 0.1+j19.7 | 0.1+j21.3 | 0.2+j38.1 | 0.3+j63.3 |
| Wire length [m] | 0.65 | 0.70 | 1.43 | 1.44 | 2.15 | 2.86 |
| Center $B_1^+$ [$\mu T/\sqrt{W}$] | 13.25 | 12.74 | 17.21 | 16.62 | 20.14 | 23.20 |

**Figure 2:** Simulated $B_1^+$ field maps and summary table showing impedance, wire length, and central $B_1^+$ amplitude for various winding configurations of a single elliptical toroid. The holder thickness was fixed at 10 mm, and each winding completes six full twists along the holder. The wire is continuous, with a single segmentation point for coil feeding. Results were normalized to 1 W accepted power.

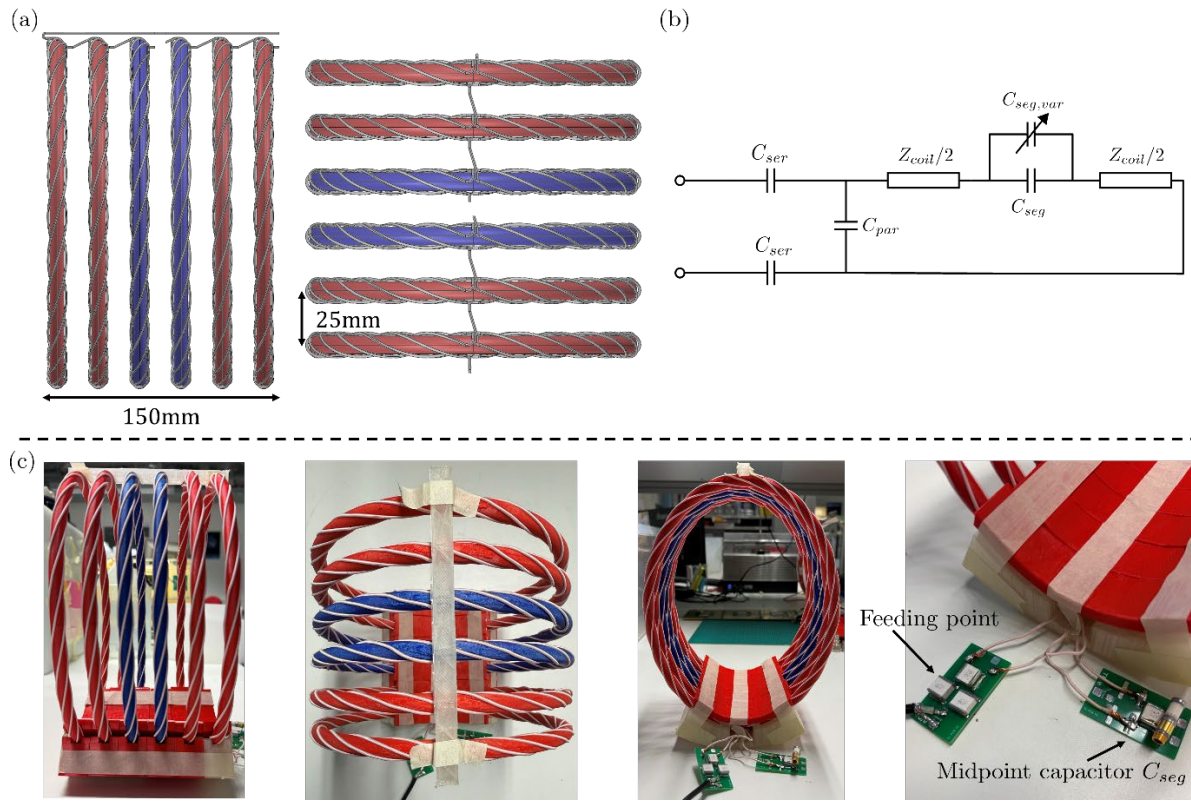

**Figure 3:** (a) Side and top views of a single full-size element with a quadruple winding, a 10 mm thick holder, and six twists per winding. (b) Schematic of the full toroidal volume coil design. (d) Photographs of the fabricated coil. The segmenting capacitor was routed externally to facilitate tuning during in vivo measurements.

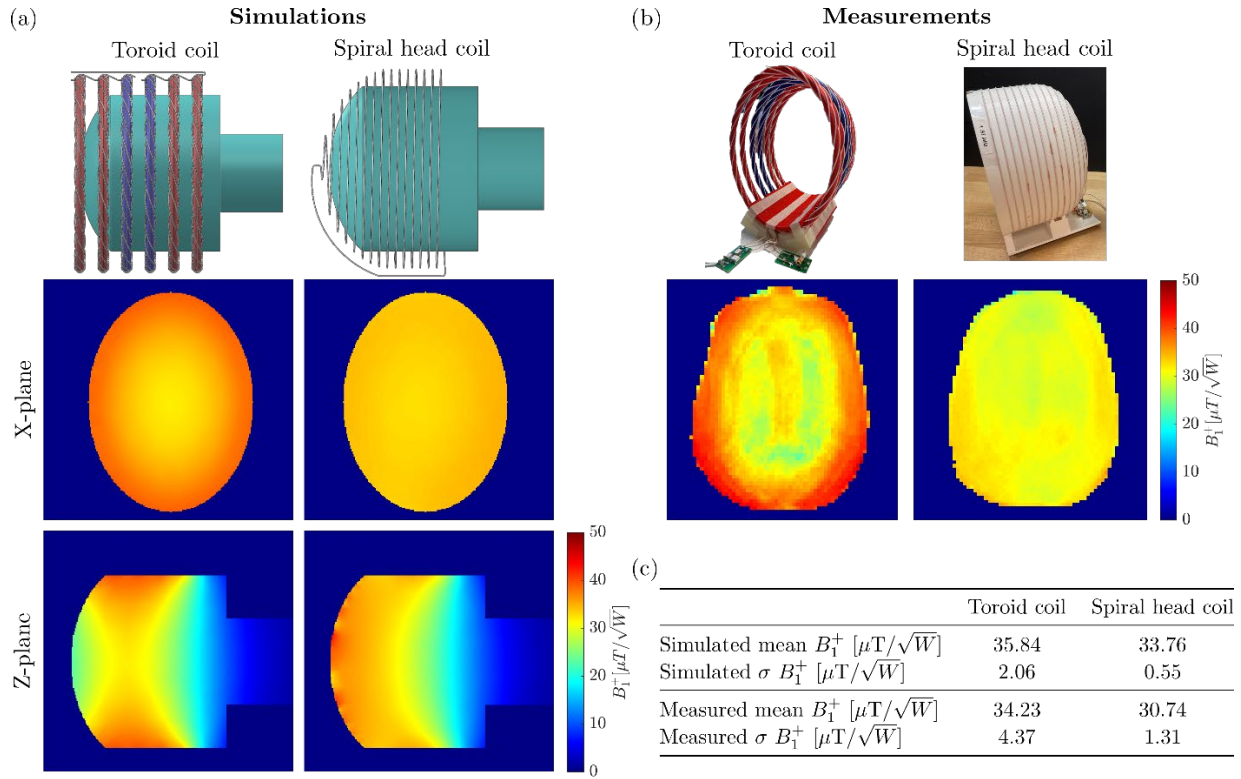

**Figure 4:** (a) Simulated $B_1^+$ field distributions for the toroidal and spiral head coils, normalized to 1 W input power. (b) Measured $B_1^+$ field maps for both coils, normalized to their respective input power. (c) Comparison of the mean and standard deviation of the $B_1^+$ field from simulations and measurements.

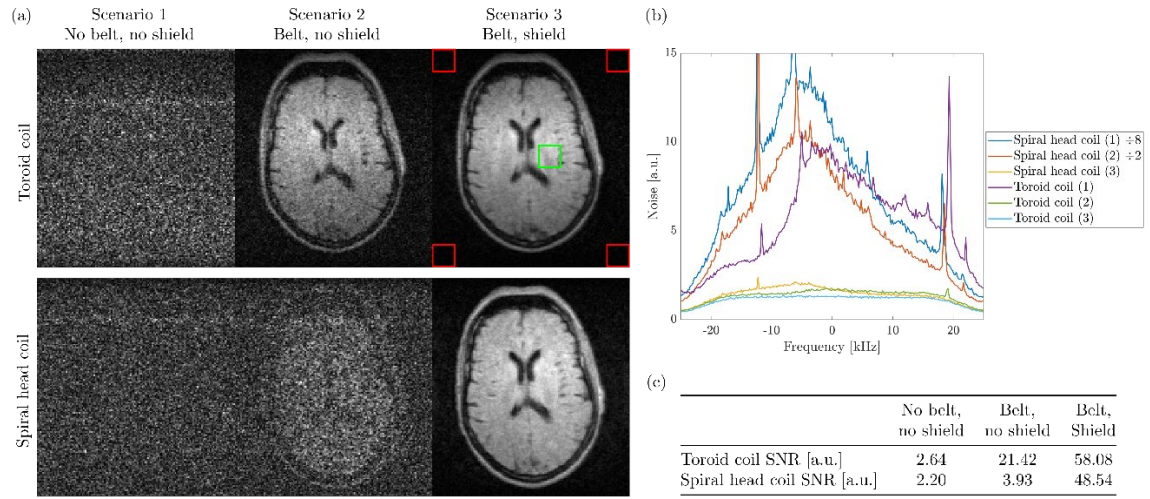

**Figure 5:** (a) In-vivo TSE images acquired with the toroidal and spiral head coils under three different grounding conditions: no shielding, with a grounding belt, and with both grounding belt and aluminum shield. (b) Measured noise peak values for both coils across the same conditions. (c) Table with calculated SNR values.

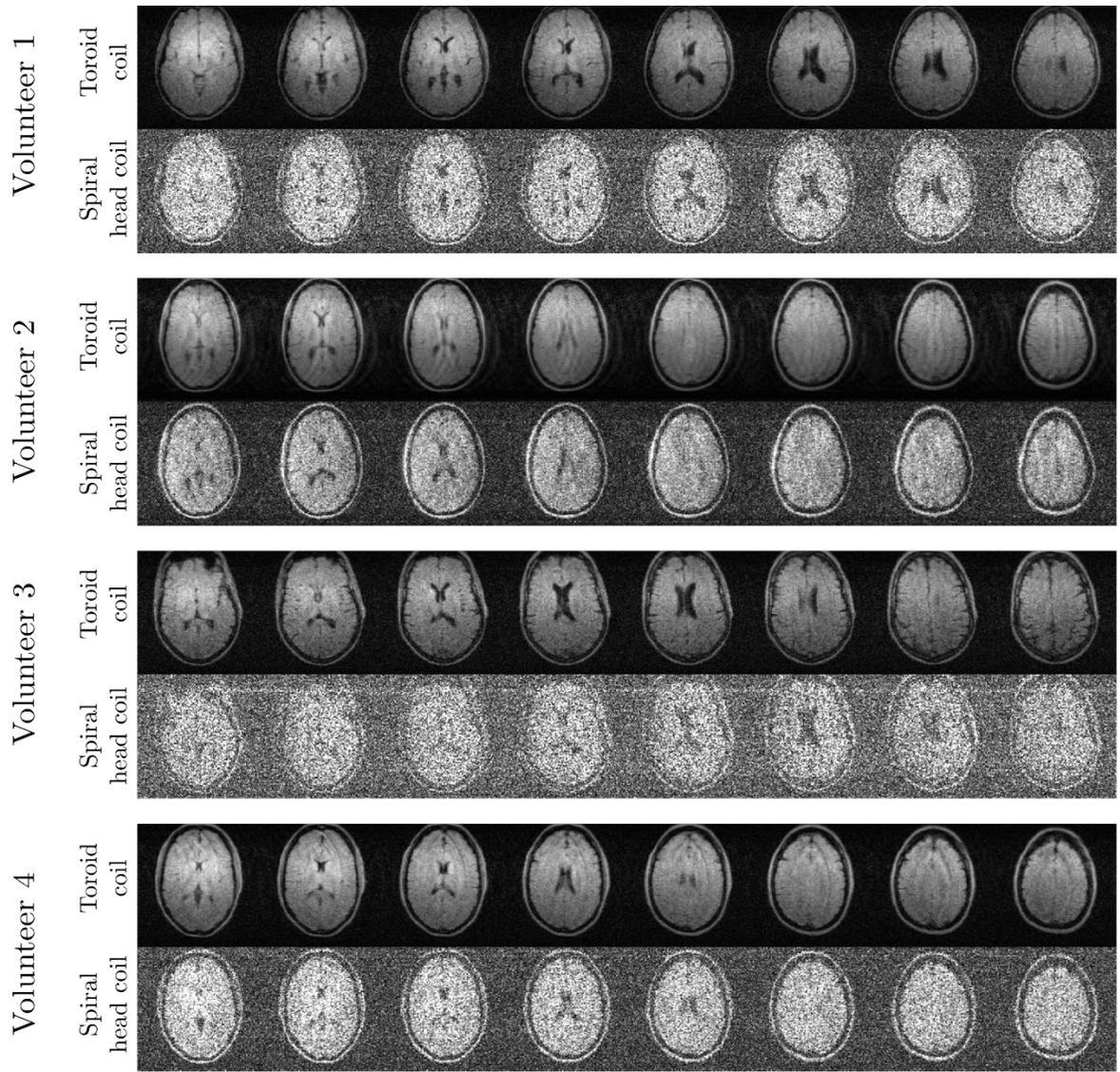

**Figure 6:** In vivo images acquired using the toroidal and spiral head coils with only a grounding belt, demonstrating reproducibility across four volunteers. Displayed are eight slices from a 3D TSE with TR/TE = 600/16 ms, FOV = 205 × 225 × 190 mm³, resolution = 1.5 × 1.5 × 5 mm³.